\documentclass[fleqn,10pt]{wlscirep}
\usepackage[utf8]{inputenc}
\usepackage[T1]{fontenc}
\usepackage{gensymb}
\usepackage{float}
\usepackage{xcolor}

\title{Enhanced laser-driven proton acceleration via improved fast electron heating in a controlled pre-plasma}

\author[1,2,@]{Leonida A. Gizzi}
\author[3,4,§]{Elisabetta Boella}
\author[1,2,*]{Luca Labate}
\author[1]{Federica Baffigi}
\author[3]{Pablo J. Bilbao}
\author[1]{Fernando Brandi}
\author[1]{Gabriele Cristoforetti}
\author[5,6]{Alberto Fazzi}
\author[1]{Lorenzo Fulgentini}
\author[7]{Dario Giove}
\author[1]{Petra Koester}
\author[1]{Daniele Palla}
\author[1]{Paolo Tomassini}

\affil[1]{Intense Laser Irradiation Laboratory, INO-CNR, Pisa, Italy}
\affil[2]{INFN, Sez. Pisa, Italy}
\affil[3]{Physics Department, Lancaster University, Bailrigg, Lancaster LA1 4YW, UK}
\affil[4]{Cockcroft Institute, Sci-Tech Daresbury, Keckwick Lane, Warrington WA4 4AD, UK}
\affil[5]{Dipartimento di Energia, Politecnico di Milano, Italy}
\affil[6]{INFN, Sezione di Milano, Italy}
\affil[7]{INFN-LASA, Segrate, Italy}
\affil[@]{correspondence to leonidaantonio.gizzi@cnr.it}
\affil[§]{e.boella@lancaster.ac.uk}
\affil[*]{luca.labate@ino.cnr.it}

\begin{abstract}
 The interaction of ultraintense laser pulses with solids is largely affected by the plasma gradient at the vacuum-solid interface, which modifies the absorption and ultimately, controls the energy distribution function of heated electrons. A micrometer scale-length plasma has been predicted to yield a significant enhancement of the energy and weight of the fast electron population  and to play a major role in laser-driven proton acceleration with thin foils. We report on recent experimental results on proton acceleration from laser interaction with foil targets at ultra-relativistic intensities. We show a three-fold increase of the proton cut-off energy when a micrometer scale-length pre-plasma is introduced by irradiation with a low energy femtosecond pre-pulse. Our realistic numerical simulations agree with the observed gain of the proton cut-off energy and confirm the role of stochastic heating of fast electrons in the enhancement of the accelerating sheath field.
 
\end{abstract}
\begin{document}

\flushbottom
\maketitle

\thispagestyle{empty}

\section*{Introduction}

Ion acceleration driven by ultraintense lasers using Target Normal Sheath Acceleration (TNSA) \cite{Snavely} is establishing itself as a powerful technique to access relatively high energy ion beams in a compact and affordable layout. 
Since the original investigations, effort has been dedicated \cite{Daido, Macchi2013} to enhance the cut-off energy and the flux of the accelerated ions via a number of advances of laser and target specifications using plasma mirror for ultra-high laser contrast with ultra-thin targets \cite{Kaymak2019}, cryogenic \cite{Scott2018} or nano-structured targets \cite{Ferri2019}.  To date, however, practical exploitation of laser-driven ion acceleration relies heavily on the original TNSA configuration based on thin foil targets and optimized contrast without plasma mirror, possibly operating at the repetition rate required for applications like radiobiology \cite{Bayart2019, Giulietti_2016}, where high dose irradiation is needed for meaningful studies.
In this context, great attention is being dedicated to the control of accelerated ions, including energy cut-off, beam divergence, charge and emittance. Target optimisation and engineering, looking at different properties of surface, geometry and conductivity, are becoming crucial in this effort to ensure reliable and reproducible performance of beamlines based on laser-driven acceleration. 

From a more fundamental viewpoint, recent investigations to further improve ion acceleration performance focus on the interaction parameters affecting the energy density of laser-heated fast electrons, namely their divergence, peak current and temperature. It is well known that fast electron generation is strongly modified by the size of the laser focal spot \cite{Dover2020, Steinke_2020}, and more effectively by the scale-length of the plasma at the vacuum-target interface. Indeed, several modelling studies based on Particle-In-Cell (PIC) simulations predict a significant increase of the proton cut-off energy with optimized pre-plasma conditions. In \emph{Sentoku et al.}\cite{Sentoku2002}, simulations were used to investigate fast electron heating in the standing wave occurring in the pre-plasma from superposition of incident and reflected wave, showing a linear increase of the ion energy cut-off with the scale-length of the pre-plasma. Similar results were shown in a more recent modelling study \cite{Kumar2019} where the role of temporal pulse shape was also explored.  The role of standing waves was then further investigated in a colliding pulses configuration with two pulses incident on the target from opposite incident angles \cite{Ferri2019colliding}, showing enhancement in the fast electron temperature and ion energy cut-off for both ultra-high contrast laser systems using double plasma mirror. In the same study, a strong linear increase of the cut-off energy with the scale-length of the pre-plasma is shown for both colliding pulse and single pulse at constant energy, with the highest proton cut-off energy occurring for a micrometer scale-length plasma.
More in details, in \emph{Sentoku et al.} \cite{Sentoku2002}, a moderate increase of the proton energy up to 1.5 times was found at mildly relativistic intensities of  $\mathrm{10^{18}}\:\mathrm{W/cm^{2}}$ at 1.053 $\mathrm{\mu m}$ wavelength for a micrometer-scale pre-plasma. The increase was explained as due to the stochastic heating of electrons in the standing wave structure responsible for an enhanced ponderomotive acceleration of electrons. In the same study it is shown that this enhancement of proton energy can be expected to grow to a factor of 4 or more for longer scale-length and eventually saturate above 6-8 $\mathrm{\mu m}$ scale-length. On the other hand, this enhancement factor may be partially reduced at ultra-relativistic intensities due to the $\propto \sqrt{I}$ scaling compared to the $\propto I$ of moderately or sub-relativistic intensities. In \emph{Ferri et al.} \cite{Ferri2019colliding} simulations at a higher intensity of $\mathrm{7 \times 10^{19}}\:\mathrm{W/cm^{2}}$ at 0.8 $\mathrm{\mu m}$ wavelength ($\mathrm{\mathrm{a_{0} = 5.7 }}$) leads approximately to a three-fold increase of the proton cut-off for single pulse irradiation of a 3 $\mathrm{\mu m}$ thick Al target and for a density scale-length of 0.6 $\mathrm{\mu m}$.
In \emph{Kumar et al.} \cite{Kumar2019} simulations at ultra-relativistic intensities of $\mathrm{10^{20}}\:\mathrm{W/cm^{2}}$ and $\mathrm{\mathrm{a_{0}=10.6}}$, very similar to the value of our experiment, were investigated. In this intensity regime PIC simulations show that in the presence of a pre-plasma consisting of a 10 $\mathrm{\mu m}$ linear density ramp, the proton cut-off energy is expected to be approximately 8.5 MeV, while without the pre-plasma, the cut-off energy is 5 MeV, showing again an enhancement of approximately a factor 1.7.

In spite of these theoretical predictions, to date, a clear experimental evidence of stochastic heating of fast electrons in proton acceleration measurements is still lacking.  This is mainly due to  the complex interplay typically occurring in real laser-target interactions, between plasma formation at the front side and the deformation and disruption of the target rear side, where ion acceleration mainly occurs, making changes to the fast electron distribution function impossible or difficult to detect. In fact, in most previous experiments, the generation of the required plasma gradient at the vacuum-solid interface is achieved with uncontrolled precursor laser irradiation due to Amplified Spontaneous Emission (ASE) \cite{Kaluza2004} or nanosecond irradiation \cite{McKenna2008} that have unpredictable, detrimental effects on the target integrity.

The use of low energy, femtosecond pre-pulse for plasma generation allows a controlled and predictable pre-plasma formation (see Methods). Previous attempts to use this technique to show the effect of pre-plasma on fast electron heating, were hindered by the relatively low intensity \cite{Yogo2007} or by the presence of the ASE \cite{Glinec2008}. More recently \cite{Wang2019}, the same technique was used with ASE-free interaction at ultra-relativistic intensities. A modest increase of the proton cut-off energy was shown for 2 $\mu$m thick plastic targets with numerical simulations suggesting self-focusing and pulse-front steepening.

We consider a similar framework to investigate ultra-relativistic interaction with thin foil targets characterized by a vacuum-solid interface consisting of a micrometer scale-length pre-plasma (see Methods). The interaction configuration is modelled using realistic numerical PIC simulations that account for the presence of the pre-plasma at the vacuum-target interface. Simulations point at the presence of a strong stochastic heating of fast electrons which is responsible for a robust enhancement of the TNSA field that boosts the proton energy. Our measurements confirm this scenario, showing an unprecedented three-fold increase of the cut-off energy of accelerated protons compared to the measurements obtained without pre-plasma.  While unfolding the effect of a novel physical mechanism, our results also provide a recipe for the optimum target thickness for a given pre-pulse timing, opening a novel route to the optimization and control of efficient and robust TNSA ion sources.

\section*{Results}
\begin{figure}
\begin{center}
\includegraphics[height=6cm]{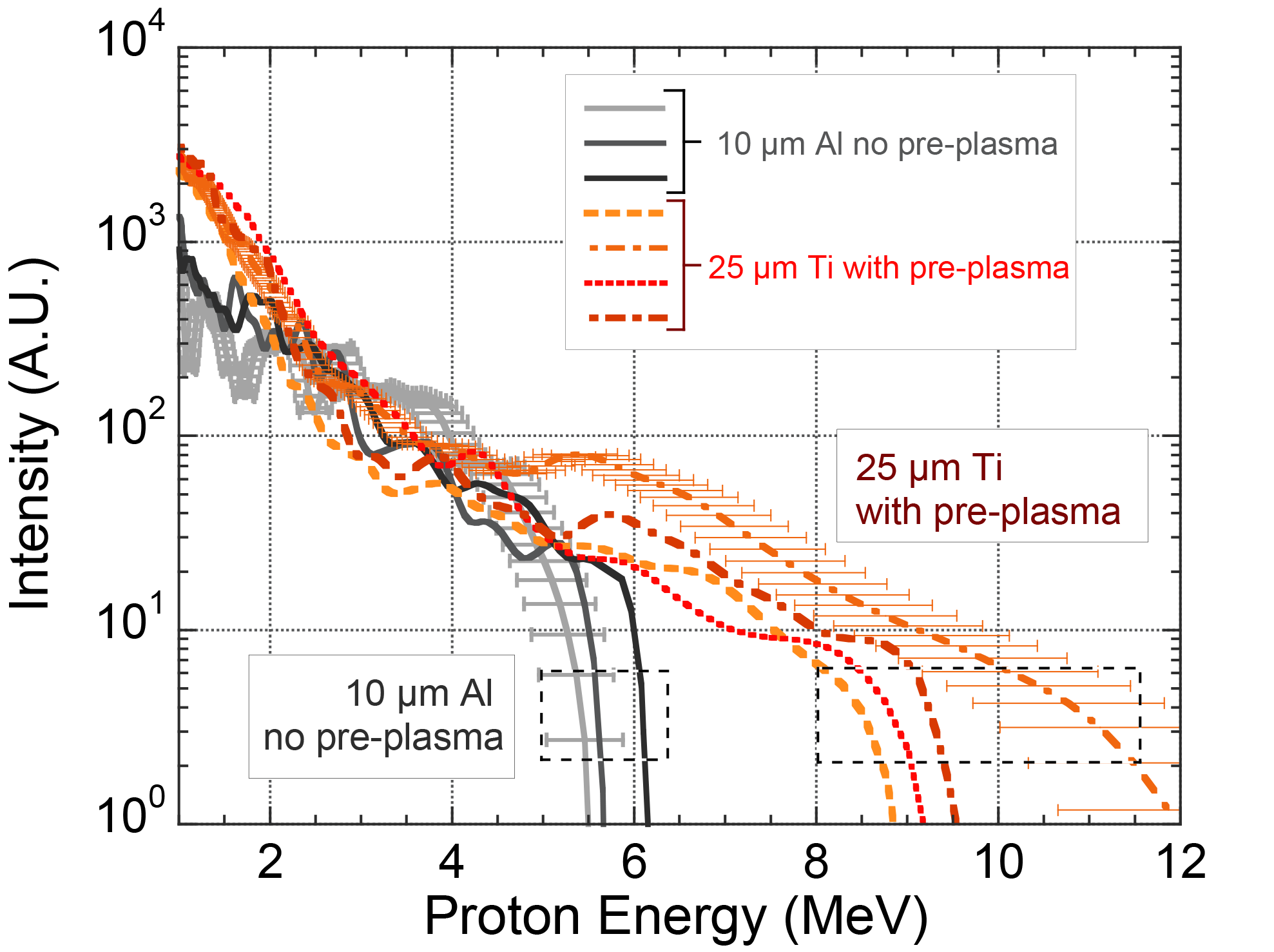}
\includegraphics[height=6cm]{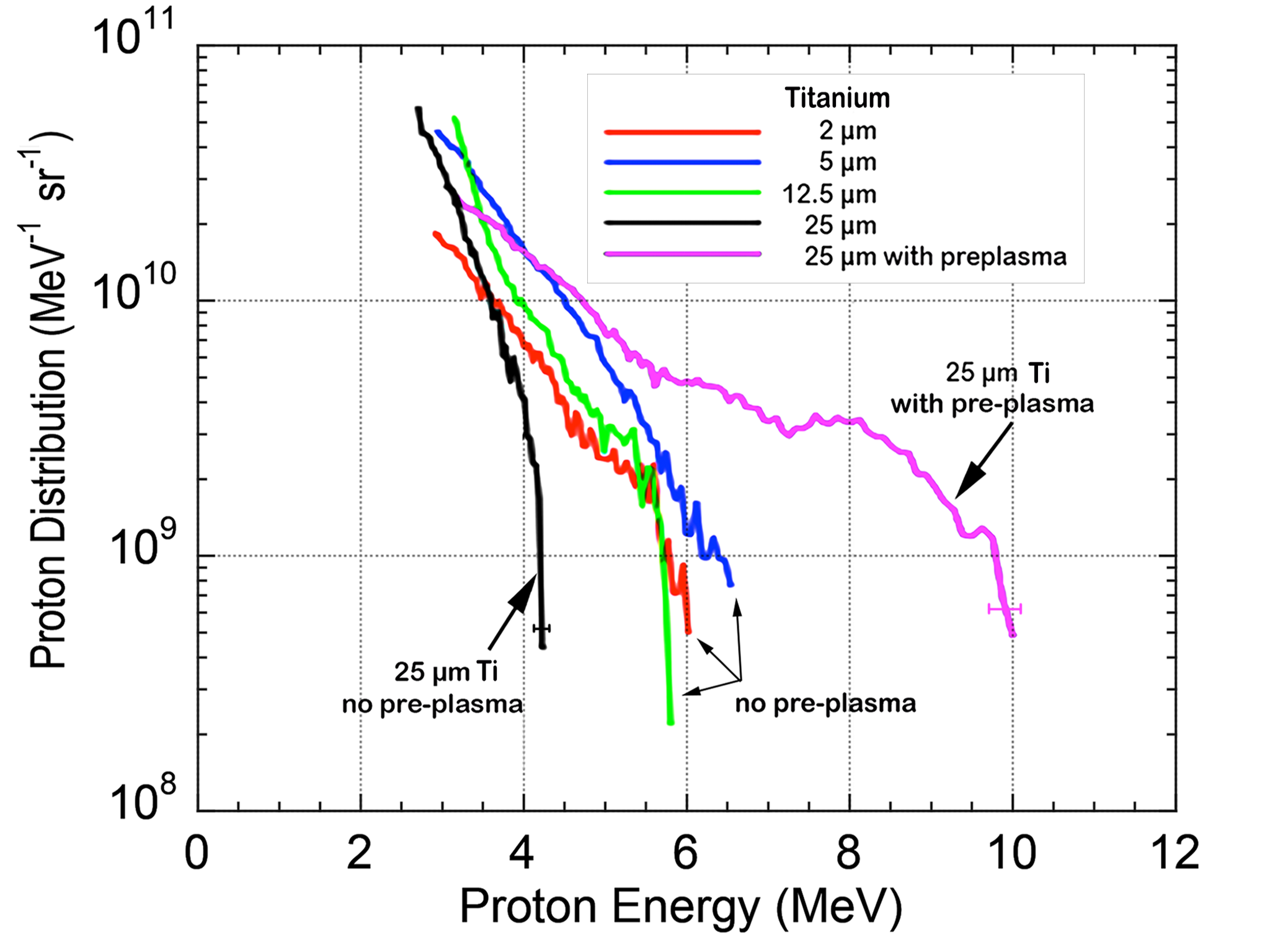}
\caption{(left) Comparison of experimental proton spectra obtained from the TPS for 10 $\mathrm{\mu m}$ thick Al targets without pre-plasma and 25 $\mathrm{\mu m}$ thick Ti target with the pre-plasma. Sample error bars on the measured proton energy are shown for two spectra and are indicative of the error-bars for all the curves.  (right) Comparison of experimental spectra obtained from deconvolution of the temporal evolution of the TOF signals from a set of Titanium foil targets of different thickness of 2, 5, 12.5 and 25 $\mathrm{\mu m}$. All data were taken without the pre-plasma, except the purple line for 25 $\mathrm{\mu m}$ with cut-off energy at 10 MeV. Sample error bars on the measured proton energy are shown on the two spectra with 25 $\mu$m Ti targets. \label{fig:exp_data}}
\end{center}
\end{figure}

\textbf{Experiment.}   The experiment was carried out as part of the development of a laser driven particle acceleration programme for the development of a proton beamline \cite{Gizzi2018_2}. Details of the full experimental set up are given in the Methods and a full characterization of the interaction conditions is given elsewhere \cite{Gizzi2018_2,Gizzi2020,Cristoforetti2020}. Here we focus on a set of proton spectra obtained from irradiation of thin foil targets at a peak intensity of $\mathrm{2.4\times10^{20}}\:\mathrm{W/cm^{2}}$, with and without a pre-plasma forming fs pre-pulse. Proton spectra were obtained simultaneously with two different detectors. The plots of Fig.\ref{fig:exp_data}(left) shows two sets of proton spectra obtained with the Thomson Parabola Spectrometer (TPS) (see Methods) on 10 $\mathrm{\mu m}$ Al targets and on 25 $\mathrm{\mu m}$ Ti targets, respectively. The three spectra with 10 $\mathrm{\mu m}$ Al targets were obtained \textit{without} the pre-plasma and with the standard contrast of our laser\cite{Gizzi2018_2}, and show a cut-off energy around 6 MeV. The four spectra with 25 $\mathrm{\mu m}$ Ti targets were obtained \textit{with} a  pre-plasma generated by the fs pre-pulse arriving 10.4 ns before the main pulse, having a scale-length of 4.5 $\mu$m at the relativistic critical density at the time of interaction (see Methods).

The proton spectra of Fig.\ref{fig:exp_data} (left) obtained with the 25 $\mathrm{\mu m}$ Ti, in the presence of the pre-plasma,   show a cut-off energy in the range between $\sim$9 and 12 MeV, namely up to a 2-fold increase with respect to the 10 $\mathrm{\mu m}$ Al target. These results clearly show that, in spite of the significant increase of target thickness compared to the 10 $\mathrm{\mu m}$ Al, the presence of the pre-plasma significantly modifies the proton spectrum obtained with the 25 $\mathrm{\mu m}$ Ti, with a strong enhancement of the maximum energy of accelerated protons. A systematic increase of the signal at low energy is also visible for the Ti data compared with Al targets. The 25 $\mathrm{\mu m}$ Ti plots of Fig.\ref{fig:exp_data}(left) also demonstrate that the observed increase of the cut-off energy in the presence of a pre-plasma is repeatable and statistically significant.

To rule out the the possibility that this difference could be explained by the different target material, shots were taken with Ti targets of different thickness with and without the pre-plasma. Fig.\ref{fig:exp_data} (right) shows the proton spectra obtained with the Time Of Flight (TOF) detector (see Methods) from the irradiation of Ti targets with thickness ranging from 2 to 25 $\mathrm{\mu m}$. All the data were taken without the pre-plasma, except the purple line with cut-off at 10 MeV that was obtained with the pre-plasma. According to these plots, the cut-off energy without the pre-plasma is maximum around 7 MeV for a 5 $\mathrm{\mu m}$ Ti target and drops slightly to 6 MeV for 12.5 and to 4 MeV for 25 $\mathrm{\mu m}$ Ti.

This is consistent with the expected behaviour for TNSA accelerated protons where cut-off energy is expected to decrease with increasing target thickness\cite{Daido, Macchi2013}. As already observed in our experimental conditions\cite{Gizzi2018_2}, the cut-off energy also decreases for the thinnest target of 2 $\mathrm{\mu m}$ Ti. This is expected due to the finite temporal contrast of our laser system at the ps level as previously discussed in details\cite{Gizzi2018_2}.  Similar cut-off energies were obtained for 2.4 $\mathrm{\mu m}$ Al targets that showed a cut-off energy of <4 MeV \cite{Gizzi2018_2}.  

Remarkably, Fig.\ref{fig:exp_data} (right) shows that the cut-off energy of the 25 $\mathrm{\mu m}$ Ti target with the pre-plasma exhibits the highest value of 10 MeV, a value  approximately 2.5 times the value obtained for the same target thickness without the pre-plasma. These results confirm those obtained with the TPS, as shown in Fig.\ref{fig:exp_data} (left) where a cut-off energy as high as 12 MeV is shown, corresponding to a three-fold increase of the cut-off energy without the pre-pulse for the same 25 $\mathrm{\mu m}$ Ti target thickness. 
We observe that the highest cut-off energy with the pre-plasma was found only for the 25 $\mathrm{\mu m}$ thick Titanium target. In all other explored cases of thinner targets, a lower or similar cut-off energy was found in the pre-plasma case, compared with the no pre-plasma case, when the same target thickness was used. As discussed below, this is due to the effect of the shock launched by the pre-pulse used to generate the pre-plasma that, for thinner targets, reaches the rear surface before the arrival of the main pulse, disrupting the sheath field.

\begin{figure}
\begin{center}
\includegraphics[scale=0.35]{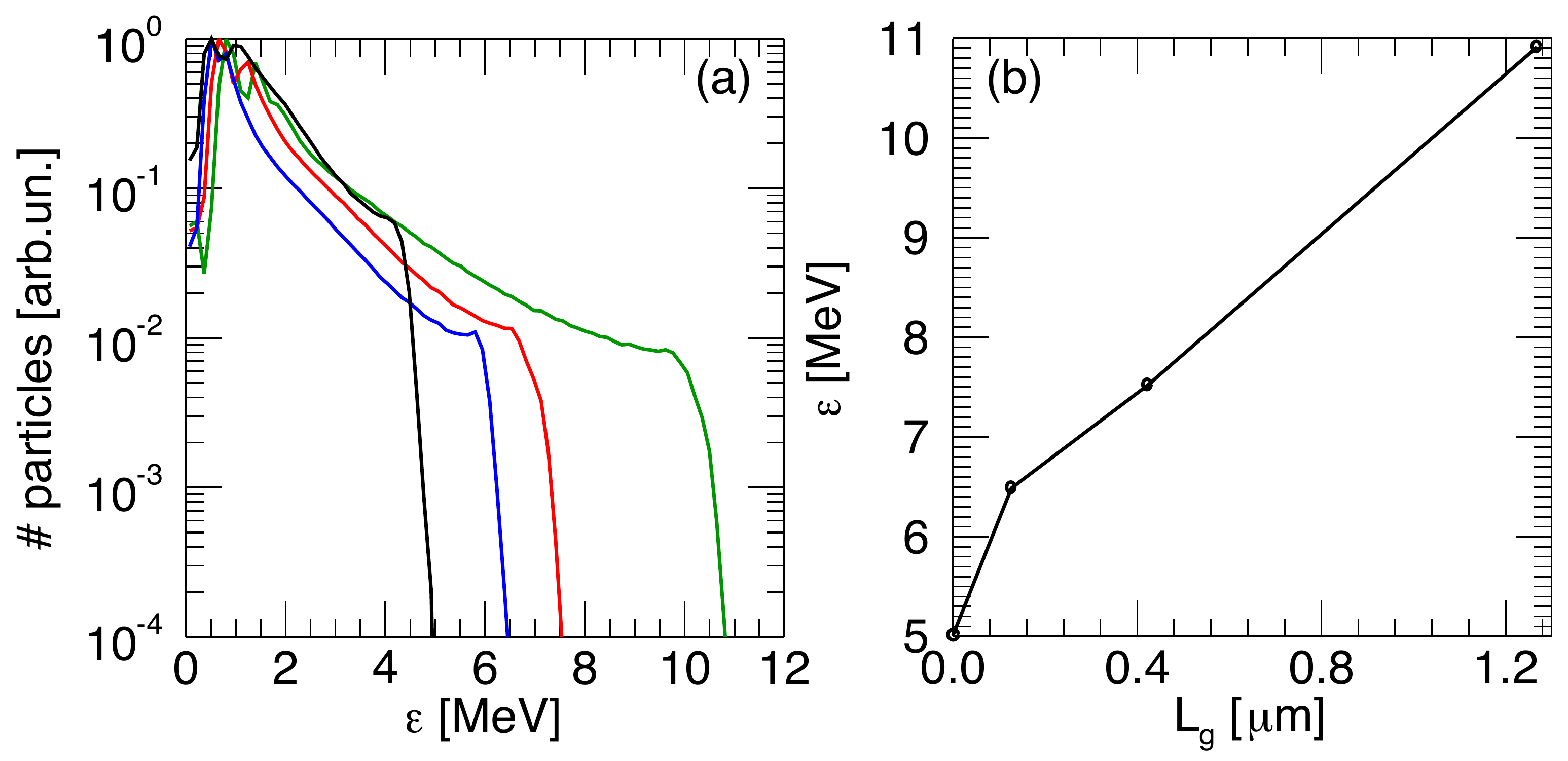}
\caption{3D PIC simulation results: (a) Proton energy spectra for a $10 \: \mathrm{\mu m}$ Al target for different pre-plasma scalength: $L_g = 0.13$ (blue), $0.42$ (red) and $1.27 \: \mathrm{\mu m}$ (green). The black curve corresponds to a sharp plasma-vacuum transition. (b) Cut-off energy vs pre-plasma scalength for the cases in (a).} 
\label{fig:simulations}
\end{center}
\end{figure}

\vspace{1mm}
\noindent
\textbf{Modelling.} We carried out extensive PIC simulations (see Methods) to model our interaction conditions and ion acceleration measurements and to identify the mechanism responsible for the observed three-fold increase of the proton cut-off energy in the presence of a pre-plasma. We considered a pre-plasma scale-length ranging from a minimum of 0.13 up to 5 $\mathrm{\mu m}$.
The spectrum of accelerated protons obtained from 3D PIC simulations for a 10 $\mathrm{\mu m}$ Al for the case of no pre-plasma and for values of the plasma scale-length of $L_g = 0.13 \: \mathrm{\mu m}$, $L_g = 0.42 \: \mathrm{\mu m}$ and $L_g = 1.27 \: \mathrm{\mu m}$ are shown in Fig.\ref{fig:simulations}(a), while Fig.\ref{fig:simulations}(b) reports the proton cut-off energy for the same cases. Here, the spectra are plotted when the proton energy has reached the saturation value in the simulation.

According to these results, simulations predict a cut-off energy just below 5 MeV for the pre-plasma free case. This value increases up to 7.5 MeV for the 420 nm plasma scale-length and exceeds 10 MeV for the 1.27 $\mathrm{\mu m}$ scale-length. These simulations show a linear dependence of the cut-off energy with the pre-plasma scale-length, predicting a cut-off energy increasing rapidly with the pre-plasma scale-length, with a two-fold increase for a scale-length of 1.27 $\mathrm{\mu m}$.

A comparison with our experimental results of Fig.\ref{fig:exp_data}(left) for the case of 10 $\mathrm{\mu m}$ Al without pre-plasma, yielding a measured cut-off energy  of 6.0 MeV, shows a similar value for a pre-plasma scale-length of approximately 100 nm.  As anticipated, this scale-length is consistent with the presence of the ps pedestal of our laser pulse. In fact, taking into account the ps contrast our laser system \cite{Gizzi2018_2}, we expect the formation of a very short scale-length pre-plasma a few ps before the laser peak. Considering a pre-plasma expanding at the ion-acoustic wave speed, we estimate a density scale-length at the critical density of the order of 100 nm, a value that is fully consistent with the scale-length expected from the numerical simulations results discussed above.
Fig.\ref{fig:simulations} confirms that in the case of a 10 $\mathrm{\mu m}$ Al target the proton cut-off energy could exceed 10 MeV already for a modest pre-plasma scale-length just above 1.27 $\mathrm{\mu m}$.  However, we stress here that in these and other similar PIC simulations, the presence of the pre-plasma in front of the target is included \textit{ad hoc}. In reality, the formation of any pre-plasma generated on the target surface by either the ps pedestal or by a custom pre-pulse, also involves the bulk of the target. The shock generated at the target surface by any pre-pulse interaction eventually reaches the rear surface of the target, disrupting the formation of the TNSA field.  This problem can only be neglected in the case of sufficiently thick targets, where shock break-out at the target rear surface occurs after the main pulse interaction, so that TNSA and ion acceleration take place before any disruption of the target rear surface has occurred.

\vspace{1mm}
\noindent
\textbf{Shock break-out.} Here we refer to the collisional shock generated by the interaction of the pre-pulse at the target surface. We point out that shock generation inside the target, driven by fast electrons, has also been considered elsewhere \cite{Shaikh2018} for ultrarelativistic intensities. In our experimental conditions, this effect can be neglected for the pre-pulse, due to the low intensity. As for the main pulse, break-out at the target rear face of any shock produced by fast electrons would anyway take place tens of picoseconds after proton acceleration \cite{Shaikh2018}, so this effect is not expected to play a role in our measurements. The role of collisional shock generation induced by pre-pulse irradiation at the target surface and propagation in the bulk target was investigated with our hydrodynamic simulations (see Methods). We found a shock wave propagation speed of 1.7 $\mathrm{\mu m}$/ns for the Al target. Consequently, in the case of the 10 $\mathrm{\mu m}$ Al target, shock break-out occurs 5.9 ns after pre-pulse irradiation, that is 4.5 ns \textit{before} main pulse irradiation that in our experiment occurs 10.4 ns after pre-pulse irradiation. 
Therefore, in our experimental conditions, the enhancement seen in PIC simulations for the 10 $\mathrm{\mu m}$ Al target with 1.27 $\mathrm{\mu m}$ scale-length could not be observed because disruption of the rear surface would occur before main pulse irradiation.

In contrast, for Ti targets the hydrodynamic code predicts a shock propagation speed of 1.5 $\mathrm{\mu m}$/ns that, for the 25 $\mathrm{\mu m}$ Ti target, gives a shock break-out time at the rear surface of 16.6 ns after pre-pulse irradiation. In our experimental conditions this  corresponds to 6.2 ns after the main pulse irradiation, thus allowing TNSA to take place without perturbation. These considerations indeed show that the presence of the pre-plasma for Al targets below 19 $\mathrm{\mu m}$ and Ti targets below approximately 17.5 $\mathrm{\mu m}$ are subject to the effect of the shock break-out at the rear face and disruption of the TNSA field. This also explains why higher cut-off energy was only found for the 25 $\mathrm{\mu m}$ thickness Ti among all explored targets.  Incidentally we observe that the TNSA proton cut-off energy for sharp vacuum-target interface is indeed expected to decrease with increasing target thickness, thus making such relatively thick targets, in general, not ideal for efficient proton/ion acceleration. On the other hand, both  simulations and experimental data clearly show that for a sufficiently large scale-length pre-plasma, the gain in cut-off energy due to interaction with the pre-plasma can largely overcome such losses and lead to record cut-off values. 
\begin{figure}[H]
\begin{center}
\includegraphics[scale=0.43]{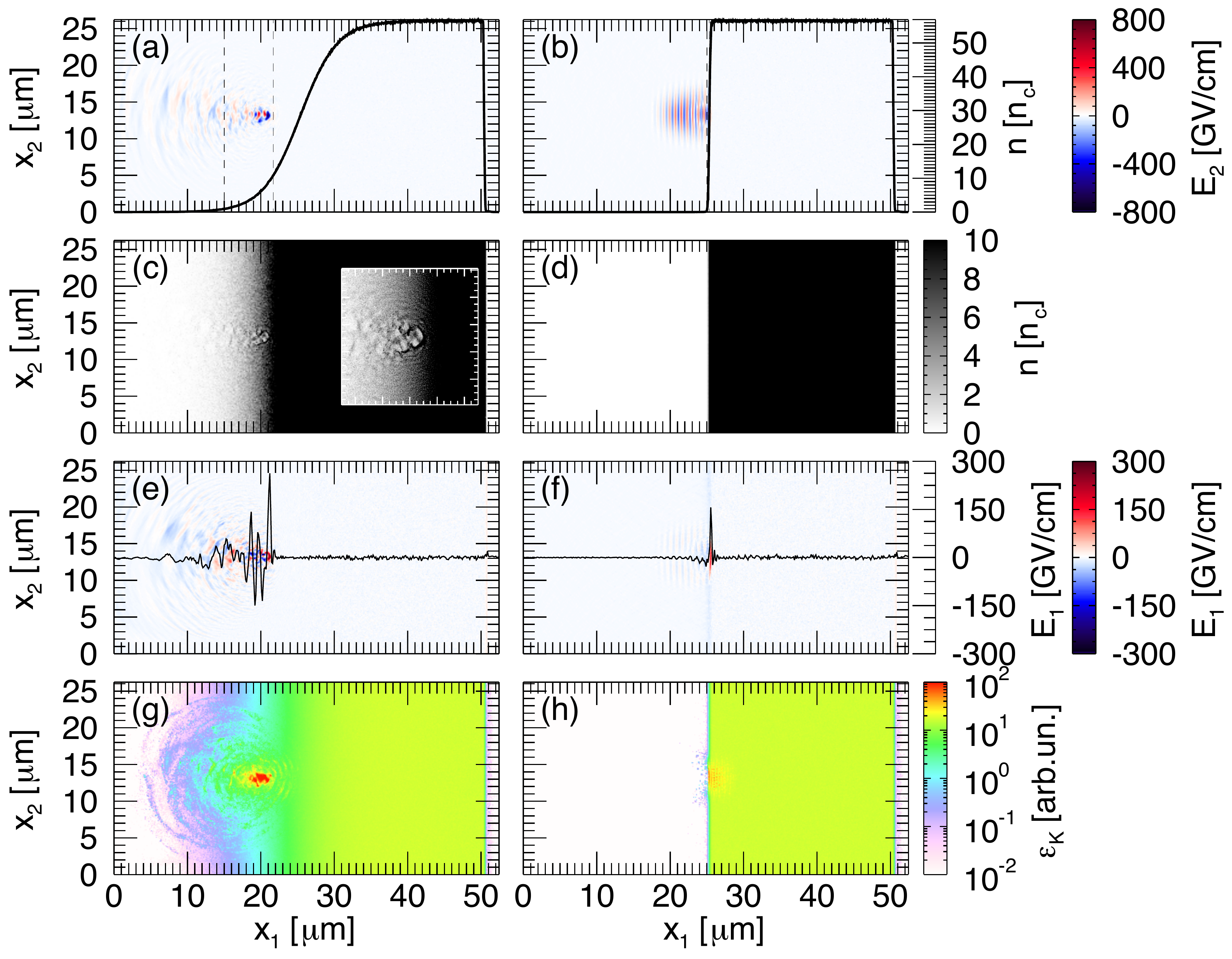}
\caption{2D PIC simulations: (a, b) Laser electric field and longitudinal density profile (black solid line), (c, d) electron density, (e, f) longitudinal electric field and (g, h) electron kinetic energy density at $t = 0.11 \: \mathrm{ps}$ for a Ti plasma with $L_g = 5$ (left column) and $0.13 \: \mathrm{\mu m}$ (right column). The black dashed lines in (a, b) denote the position of the critical density and the relativistic critical density. The inset in (c) is an enlargement showing details of the electron cavities on the target surface. The black solid lines in (e, f) represent the electric field lineout in the middle of the simulation box.}\label{fig:simulations2}
\end{center}
\end{figure}
\textbf{Scaling.} In order to investigate the interaction with a more extended pre-plasma, we also carried out 2D PIC simulations (see Methods) with a scale-length beyond the value of 1.27 $\mathrm{\mu m}$ explored above with 3D simulations. Being computationally much more affordable that the 3D simulations, 2D simulations allowed us to explore the role of longer pre-plasma gradient in front of the target with scale-lengths of the order of those predicted by hydrodynamic simulations (see Methods). 

In Fig.\ref{fig:simulations2} the results from a simulation with a long pre-plasma ($L_g = 5 \: \mathrm{\mu m}$) are compared with results obtained with $L_g = 0.13 \: \mathrm{\mu m}$ .  
When a long plasma gradient is present in front of the target, the laser penetrates through the underdense plasma (Fig.\ref{fig:simulations2}(a)), where it undergoes self-focusing and steepening. When the laser reaches the plasma region where $n_c \le n \le a_0n_c$, $n$ being the electron density, $n_c=m_e \omega^2/(4\pi e^2)$ being the critical density and $a_0 n_c$ relativistic critical density for $a_0>>1$\cite{Weng2012}, it expels electrons forming semi-circular cavities (see Fig.\ref{fig:simulations2}(c)) \cite{Wilks1992, Tsung2012}. It is interesting to notice that nothing of this is observed when the laser reaches the target surface when $L_g = 0.13  \: \mathrm{\mu m}$, corresponding to a nearly sharp, almost unperturbed, vacuum-plasma transition (see Fig.\ref{fig:simulations2}(b)). In the latter case the ponderomotive force associated with the laser is not large enough to push the target surface inward and bore a hole (Fig.\ref{fig:simulations2}(d)) and the laser is reflected back by the opaque target.

\section*{Discussion}
The complex laser dynamics in the plasma section across the critical density for the case $L_g = 5 \: \mathrm{\mu m}$ leads to the formation of a standing wave due to a superposition between the incident and reflected electromagnetic wave \cite{Sentoku2002, Kumar2019}. Here the role of the angle of incidence (15\degree \ in our case) has been discussed recently \cite{Chopineau2019} showing effective contribution of stochastic heating for relatively long scale-length and even larger angles of incidence. Signatures of the standing wave are clearly visible in Fig.\ref{fig:simulations2}(e) while such signatures are absent in Fig.\ref{fig:simulations2}(f), where the longitudinal electric field is plotted for $L_g = 5$ and $0.13 \:  \mathrm{\mu m}$, respectively. 
\begin{figure}[H]
\begin{center}
\includegraphics[scale=0.28]{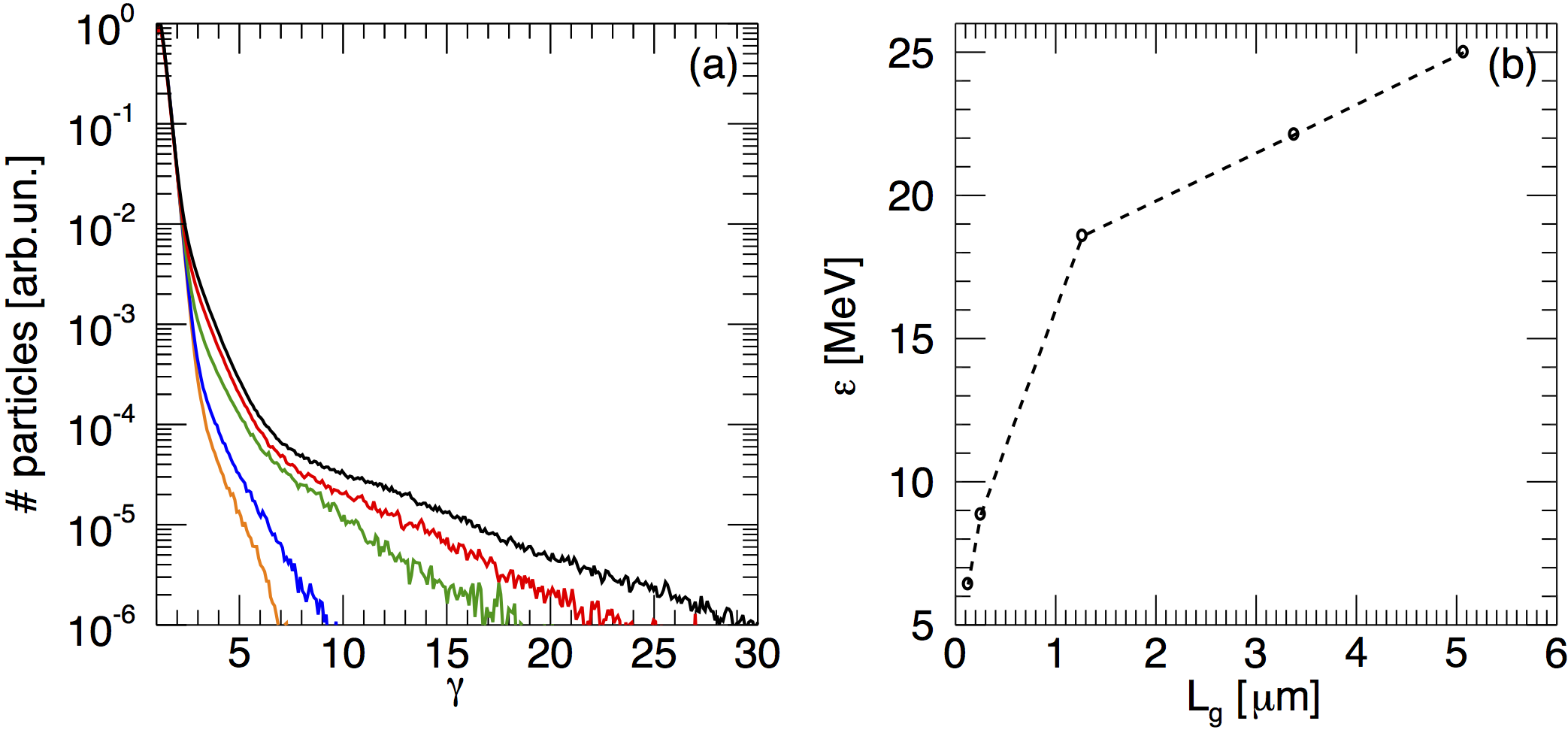}
\includegraphics[scale=0.28]{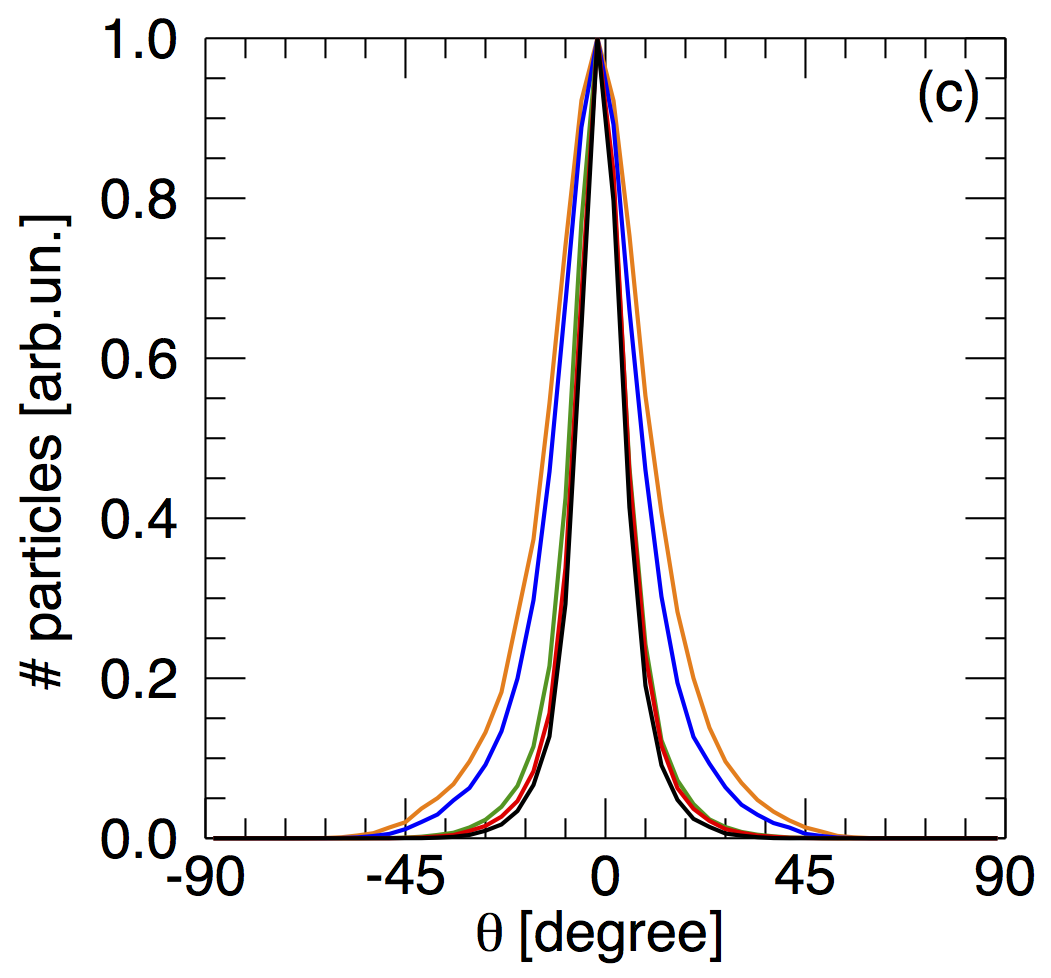}
\caption{2D PIC simulations results: (a) Electron distribution for density scale-length  $L_g = 0.13$ (orange), $0.25$ (blue), $1.27$ (green), $3.38$ (red) and $5.07 \: \mathrm{\mu m}$ (black). (b) Proton cut-off energy vs pre-plasma scale length and (c) proton angular distribution for the cases in (a).}\label{fig:simulations3}
\end{center}
\end{figure}
Electrons undergo stochastic motion in the standing wave \cite{Sentoku2002, Nuter2008} resulting in an overall more efficient electron heating, with a larger number of electrons reaching higher energies (Figs.\ref{fig:simulations2}(g) and (h)). The improved coupling between the laser pulse and the electrons in the presence of an extended pre-plasma is clearly visible in Fig.\ref{fig:simulations3}(a), which shows the electron distribution in the whole box after electrons have recirculated through the target, for different values of $L_g$. All the spectra are made up by a cold and a hot electron component. While the cold electrons have similar temperatures, the hot electrons exhibit higher temperatures for longer plasma gradients, with temperatures varying between $0.31 \: \mathrm{MeV}$ for $L_g = 0.13 \: \mathrm{\mu m}$ and $1.49 \: \mathrm{MeV}$ for $L_g = 5.07 \: \mathrm{\mu m}$. As a consequence, in the latter cases, the sheath field is enhanced and protons attain higher energies (Fig.\ref{fig:simulations3}(b)). As noted in Fig.\ref{fig:simulations}, the proton cut-off energy increases linearly with $L_g$. However, from Fig.\ref{fig:simulations3}(b) we observe that for $L_g \ge 1.27 \: \mathrm{\mu m}$, the maximum proton energy increases at a slower rate compared to shorter pre-plasma gradients. A similar behaviour was reported also in \emph{Sentoku et al.} \cite{Sentoku2002}, but for higher values of $L_g$. On one side this difference might be due to the use of a lower laser intensity in their case. On the other side it is an indication that the plasma scale-length used in our experimental conditions may already be in the saturation regime. 

A comparison with the 3D simulations of Fig.\ref{fig:simulations} shows that for the same scale-length of $1.27 \:  \mathrm{\mu m}$, 2D simulations predict a significantly higher cut-off energy of approximately 18.5 MeV, instead of 11 MeV. This is expected due to the slower decay \cite{Babaei2017} of the sheath field in 2D compared to 3D simulations, as also discussed below (see Methods). Therefore a quantitative comparison with experimental values is not straightforward. However, it is clear from these 2D simulation results
that the presence of an extended pre-plasma enables a more effective energy coupling to fast electrons that is at the origin of the higher proton cut-off energy recorded for the $25 \: \mathrm{\mu m}$ Ti target, with $L_g \ge 5 \: \mathrm{\mu m}$  and yielding the observed three-fold increase of the maximum proton energy. Finally, simulations seem to indicate that beside an increase of the cut-off energy, the presence of a longer plasma gradient leads to a more collimated proton beam (Fig.\ref{fig:simulations3}(c)). This could not be verified in our experiment and will be the subject of future investigations.

 As a final remark, we observe that further proton cut-off enhancement can be obtained optimizing the target thickness for a given pre-pulse timing relative to the main pulse and for a given plasma density scale-length. In the case of a pre-pulse to main pulse delay of 10.4 ns, as in our case, the minimum Ti target thickness to avoid the detrimental effect on shock break-out on proton acceleration is approximately 15 $\mu$m. Considering the target thickness dependence of Fig.\ref{fig:exp_data}(b), we can expect a final proton cut-off energy without pre-plasma close to the 5.8 MeV measured for the 12.5 $\mu$m target. The demonstrated three-fold increase in the presence of the pre-plasma would therefore yield approximately 17 MeV of proton cut-off energy for a 15 $\mu$m thick Ti target. Moreover, considering that, as anticipated, the 4.5 $\mu$m scale-length used in our experiment is already in the saturation regime of the fast electron heating mechanism, we can consider a shorter scale-length and a correspondingly thinner target. In the case of a 5 $\mu$m and a pre-pulse to main pulse delay of 3.3 ns, according to our model, we can expect a proton cut-off energy in excess of 20 MeV. 

Further investigation is needed to fully understand the role of laser pulse self-focusing, pulse steepening and stochastic heating of fast electrons in these interactions and their interplay. Our results demonstrate that with a reliable technique to modify the vacuum-target interface, the effect of these mechanisms on TNSA proton acceleration can be controlled to a much larger extent than  indicated by previous studies, with a potential strong impact on applications of laser accelerated proton beams.

\section*{Methods}
\noindent
\textbf{Experimental set up.}
The experiment was carried out at the Intense Laser Irradiation Laboratory using the ILIL-PW Ti:Sa laser installation \cite{gizzi_AP} based on Chirped Pulse Amplification (CPA)\cite{Strickland&Mourou_1985}. The laser pulse energy was up to 4.3 J on target with 27 fs pulse duration, with $\lambda = 0.795 \: \mathrm{\mu m}$ central laser wavelength. The 100 mm diameter beam was focused by an F/4.5 Off-Axis Parabolic (OAP) mirror with an angle of incidence of 15\degree. The focal spot was elliptical, with an average diameter of 4.4 $\mathrm{\mu m}$ (FWHM). The nominal intensity on target was up to $\mathrm{2.4\times10^{20}}\:\mathrm{W/cm^{2}}$ with a relativistic dimensionless parameter $\mathrm{\mathrm{a_{0} = 8.55 \times 10^{-10} \lambda[{\mathrm{\mu m}}]I[W/cm^2]^{1/2}} =10.6}$ The target consisted of a thin foil mounted on a remotely controlled motorized support with a sub-micrometer positioning precision. Titanium and Aluminium foils were used, with thickness ranging from 2 to 25 $\mathrm{\mu m}$. 

The main laser features and previous ion acceleration measurements are described elsewhere \cite{Gizzi_2013_AS, Gizzi2018_2, Gizzi2020_HPLSE}, including details of the experimental set up, laser contrast features and ion detectors.

Ion acceleration data were taken using the same set of ion beam diagnostics already presented in \cite{Gizzi2018_2}, including a TPS and a TOF SiC detector. The SiC detector was placed at a distance of 136 cm from the target rear side at 1.4 deg from the target normal and was filtered using a 48 $\mathrm{\mu m}$ thick Al foil. The TPS and the TOF detector were used simultaneously so that a cross-comparison of the signals obtained from the two devices was possible \cite{Gizzi1,altana}.  We observe that since the two detectors are not exactly on the same line of sight, they are sampling slightly different portions of the accelerated proton beam and, therefore, details of the spectra taken by the two detectors may be different. We refer to the Supplementary Material for additional information on the comparison between the two detectors.

In the experimental campaign described here we investigated the role of a micrometer scale-length preplasma at the vacuum-solid interface.
An effective approach to plasma formation at the target surface is based on the use of a low energy femtosecond pre-pulse \cite{Glinec2008}, provided no effect of the ASE is present. The plasma produced in these conditions is reproducible and an accurate prediction of the density scale-length at the relevant electron density and at the time of the main pulse irradiation is obtained with hydrodynamic modelling. The shock generated by the this fs pre-pulse is also be modelled and the break-out time of the shock front at the target rear face is predicted with high accuracy. This is a crucial issue in proton acceleration measurements: if the main pulse interaction occurs before the shock break-out, proton acceleration measurements will not be affected by the pre-pulse, and the role of the pre-plasma on the fast electrons can be thus identified (see Supplemental information).

In our experiment, a femtosecond pre-pulse was set up on purpose by introducing a small misalignment of the Pockels cell at the exit of the regenerative amplifier. In this way, a small fraction of the pulse circulating in the regenerative amplifier leaks through the Pockels cell every 10.4 ns. The first pulse immediately before the main pulse is partially amplified in the amplifier chain and, in particular, in the 3rd multipass amplifier, thanks to the pumping timing scheme of the entire system, while all the other pulses will be further attenuated. 
In such a way, a pre-pulse intensity on target, relative to the  main pulse, of $3\times 10^{-4}$ was obtained. The duration of this pre-pulse was measured to be of $\sim 31 \,\mathrm{fs}$ (this value was taken into account for the hydrodynamic simulations). The intensity of the pulse leaving the regenerative cavity one round-trip before the pre-pulse, that is, $20.8\,\mathrm{ns}$ before the main pulse, was measured to be smaller by at least a factor $10^{-6}$.

The effect of this pre-pulse on the target was modelled using hydrodynamic simulations to calculate the expected profile of the pre-plasma in the underdense region, at the time of interaction of the main pulse (see below). 

\vspace{1mm}
\noindent
\textbf{Numerical simulations: Particle-in-cell.}
Numerical PIC simulations were carried out using the code OSIRIS \cite{Osiris}. 
In the first set of 3D PIC simulations a 10 $\mathrm{\mu m}$ Al foil was considered with a 32 nm thick H$^+$ contaminant layer at the back of the foil. The Al foil was assumed to be pre-ionised to Al$^{3+}$ with an initial temperature of 100 keV. We verified that different level of target ionization and different initial electron temperature in the range $10 \, \mathrm{-} \, 200 \:\mathrm{KeV}$ did not lead to significant changes on the final H$^+$ energy. The density of the Al layer was 34.14 times the critical density, n$_c$ = 1.76 $\times$ 10$^{21}$ cm$^{-3}$, while the density of the H$^+$ layer on the rear side was 1.14 times the critical density. The front of the target was modelled following a hyperbolic tangent density profile centered at $x = 25.3 \: \mathrm{\mu m}$ with scale-length $L_g$ to mimic the presence of the pre-plasma.   The laser intensity used for this first set of simulations was  $\mathrm{1.6\times10^{20}}\:\mathrm{W/cm^{2}}$, corresponding to a normalized vector potential $a_0 = 8.6$. The p-polarised pulse was focused on target to a spot size $w_0$ of 2.2 $\mathrm{\mu m}$ and had a duration at FWHM of 27 fs. The simulation box was 200 $\times$ 18 $\times$ 18 $\mathrm{\mu m}^3$ with a cell size of 60 $\times$ 60 $\times$ 60 nm$^3$, equivalent to approximately 13 points per laser wavelength, with 8 particles per cell. Simulations performed with different resolution and number of particles per cell yielded similar results.

A second set of 2D PIC simulations was performed to carry out a parametric investigation of the role of target thickness and materials. For this set of simulations we employed a laser intensity on target $I = 2.4 \times \mathrm{10^{20}}\:\mathrm{W/cm^{2}} $ ($\mathrm{\mathrm{a_{0}=10.6}}$) and we modelled the interaction of this pulse with a 25 $\mathrm{\mu m}$ Ti target having a peak density $n_{Ti} = 56.9 \: n_c$ and a longitudinal profile similar to that employed in the previous simulation campaign (all the rest of the parameters are kept the same). A simulation box of 607 $\times$ 26 $\mathrm{\mu m}^2$ with a cell size of 30 $\times$ 30 nm$^2$ equivalent to about 26 points per laser wavelength has been used together with 36 particles per cell. We stress that, as it is well known \cite{Sgattoni2012, Babaei2017}, 2D simulations overestimate the final ion energy  due to the logarithmic growth in time of the proton cutoff energy. In our simulations the cutoff energies shown in Fig.\ref{fig:simulations3}b were measured at 1 ps after the beginning of the laser interaction with the target. While a one-to-one comparison with experimental results not possible, 2D numerical simulations still offer a valid insight of the underpinning physics \cite{Sgattoni2012}. 

\vspace{1mm}
\noindent
\textbf{Numerical simulations: Pre-plasma.} For the modelling of the pre-pulse generated pre-plasma, we used the 2D Eulerian hydrocode POLLUX \cite{Pert_1989}. The code models laser absorption via inverse bremsstrahlung and thermal transport via flux-limited Spitzer-Harm conductivity. Ionization is calculated assuming local thermodynamic equilibrium, while a perfect gas equation of state is used for electrons. 

According to the simulations, at the time of pre-pulse interaction with the target, the plasma has a temperature of a few eV that rapidly cools down while expanding. At the time on main pulse interaction, the scale-length of the pre-plasma is 7 $\mathrm{\mu m}$ at the classical critical density (1.7 $\times 10^{21}$ cm$^{-3}$) for the 800 nm laser light and 4.5 $\mathrm{\mu m}$ at the relativistic critical density (1.8 $\times 10^{22}$ cm$^{-3}$). Hydrodynamic simulations also show that the shock launched by the fs pre-pulse propagates through the target at a velocity of 1.5 $\mu$m/ns in the case of Titanium.

We refer to the Supplementary material for additional information on these simulations.

\section*{Conclusions}
Our study demonstrates experimentally that a micrometer-size scale-length pre-plasma can strongly enhance proton acceleration  driven by ultraintense laser interaction with not-so-thin foil targets, leading to a three-fold increase of the cut-off energy with respect to case with no pre-plasma. For targets sufficiently thick to prevent disruption of the TNSA accelerating field by shock break-out at the rear side, the observed enhancement largely overcomes the reduction of the cut-off energy due to the increased thickness.  This behaviour was modelled by detailed PIC simulations and confirmed by our measurements. Our results open the possibility to significantly extend the cut-off energy of robust TNSA based proton sources without plasma mirror or ultra-thin targets and with a minor adjustment of a standard CPA laser front-end.

\bibliography{Mendeley, references}

\section{Acknowledgements}
This project has received funding from the CNR funded Italian research Network ELI-Italy ((D.M.  No.631  08.08.2016) and from the L3IA INFN Experiment of CSN5. We gratefully acknowledge support from CLF, STFC (UK) for in kind contribution to the experimental set up described in this experiment. E.B. acknowledges the OSIRIS Consortium, consisting of UCLA and IST, for the use of the OSIRIS 4.0 framework and the visXD framework. Simulations were performed on the supercomputers ARCHER (EPCC, UK) under Plasma HEC Consortium EPSRC grants EP/L000237/1 and EP/R029148/1 and Marconi-Broadwell (CINECA, Italy) under PRACE allocation.

\section*{Author contributions statement}

L.A.G and L.L. conceived the experiment,  L.A.G, L.L., F.Ba., F.Br., G.C., A.F., L.F., D.G., P.K. and D.P. conducted the experiment, L.A.G., L.L., F.Ba., F.Br., G.C., A. F., D.G., P.K., D.P., and P.T. analysed the results. E.B. and P.J.B. carried out the numerical PIC simulations, L.L. carried out hydrodynamic simulations,  L.A.G. and E.B. prepared the manuscript. All authors reviewed the manuscript.

\end{document}